# Rosetta Brains: A Strategy for Molecularly-Annotated Connectomics

Adam H. Marblestone,[1,2] Evan Daugharthy,[2,3,4] Reza Kalhor,[2,3] Ian Peikon,[9,10] Justus Kebschull,[9,10] Seth Shipman,[2,3] Yuriy Mishchenko,[5]

Jehyuk Lee,[2,3,4] Konrad P. Kording,[6,7] Edward S. Boyden,[8] Anthony M. Zador[9] and George M. Church[1,2,3]

1 Biophysics Program, Harvard Univ., Boston, MA, USA   2 Wyss Institute for Biologically Inspired Engineering at Harvard Univ., Boston, MA, USA   3 Dept. of Genetics, Harvard Medical School, Boston, MA, USA   4 Dept. of Systems Biology, Harvard Medical School, Boston, MA, USA   5 Dept. of Engineering, Toros University, Mersin, Turkey   6 Depts. of Physical Medicine and Rehabilitation and of Physiology, Northwestern Univ. Feinberg School of Medicine, Chicago, IL 60611, USA   7 Sensory Motor Performance Program, The Rehabilitation Institute of Chicago, Chicago, IL 60611, USA   8 Depts. of Brain and Cognitive Sciences and of Biological Engineering, MIT , Cambridge, MA, USA   9 Cold Spring Harbor Laboratory, Cold Spring Harbor, NY, USA   10 Watson School of Biological Sciences, Cold Spring Harbor, NY, USA

Correspondence to: adam.h.marblestone (at) gmail.com

### Abstract

We propose a neural connectomics strategy called Fluorescent In-Situ Sequencing of Barcoded Individual Neuronal Connections (FISSEQ-BOINC), leveraging fluorescent in situ nucleic acid sequencing in fixed tissue (FISSEQ) [1, 2]. FISSEQ-BOINC exhibits different properties from BOINC [3, 4], which relies on bulk nucleic acid sequencing. FISSEQ-BOINC could become a scalable approach for mapping whole-mammalian-brain connectomes with rich molecular annotations.

## 1   INTRODUCTION

Scaling connectomics to whole mammalian brains is a challenge: the mouse brain has roughly $7.5 \times 10^7$ neurons and $> 10^{11}$ synapses in a volume of $420\ mm^3$, with kilometers of neuronal wiring passing through any cubic millimeter of tissue, and relevant anatomical features on the scale of $< 100\ nm$ [5]. We recently analyzed the design space for connectomics by studying the cost and scaling constraints on electron microscopy circuit tracing (EM) and bulk DNA sequencing of cell-identifying DNA barcode tag-pairs (BOINC) [3–5]. We also suggested using optical microscopy to map connectomes [5, 6].

Here we elaborate on the potential of the optical approach, proposing a strategy called Fluorescent In-Situ Sequencing of Barcoded Individual Neuronal Connections (FISSEQ-BOINC), leveraging fluorescent in-situ nucleic acid sequencing (FISSEQ) [1, 2]. FISSEQ-BOINC could determine the synaptic connectivity matrix, soma positions, and synapse positions, as well as diverse molecular annotations for cells and synapses.

In **Section 2**, we describe FISSEQ and propose FISSEQ-BOINC. In **Section 4**, we detail preliminary specifications for "Rosetta Brain" datasets – comprising joint, co-registered measurements of many cellular and molecular properties of a single brain – and explain how FISSEQ-BOINC could potentially meet them.

## 2   IN SITU SEQUENCING OF CO-LOCALIZED BARCODES AT SYNAPSES

Fluorescent in situ sequencing (FISSEQ) [1, 2] is a method for sequencing DNA or RNA molecules via fluorescent microscopy, in the context of intact, fixed tissue slices. In FISSEQ, a series of biochemical processing steps, such as DNA ligations or single-base DNA polymerase extensions, are performed on a block of fixed tissue, interlaced with fluorescent imaging steps. Here we illustrate the case in which DNA polymerase extension is used – this is referred to as "sequencing by synthesis", because a copy of the *sequenced* strand is *synthesized* by the polymerase. The process is *conceptually* identical to the mechanism of fluorescent sequencing by synthesis in a commercial bulk DNA sequencing machine [7], except that it is performed in fixed tissue.

Each DNA or RNA molecule in the sample is first "amplified" (i.e., copied) in-situ [8] via rolling-circle amplification [9] to create a localized "rolling circle colony" (rolony) consisting of identical copies of the parent[1] molecule [10]. A series of biochemical steps is then carried out. In the $k^{th}$ cycle, a fluorescent tag is introduced, the color of which corresponds to the identity of the $k^{th}$ base along the rolony's parent DNA strand. The system is then "paused" in this state for imaging. The entire sample can be imaged in each cycle. The fluorescent tags are then cleaved and washed away, and the next cycle is initiated. Each rolony – corresponding to a single "parent" DNA or RNA molecule in the tissue – thus appears, across a series of fluorescent images, as a localized "spot" with a sequence of colors corresponding to the nucleotide sequence of the parent molecule. The nucleotide sequence of each DNA or RNA molecule is thus read out in-situ via fluorescent microscopy.

---

[1]In the case of RNA FISSEQ, a reverse transcriptase first creates a cDNA copy, which is then circularized and amplified to generate a local DNA rolony.

The net result of this process is a form of fluorescent microscopy in which there are $4^N$ distinguishable "colors" or "labels", corresponding to the $4^N$ possible nucleotide sequences of a DNA molecule of length $N$ nucleotides. Indeed, the FISSEQ-BOINC strategy for connectomics, presented below, can be roughly conceived as a "$4^N$-color synaptic BrainBow" [5, 11–15], where $N$ is the number of bases sequenced.

**FISSEQ-BOINC**  In FISSEQ-BOINC, cell-identifying RNA barcodes [3, 4, 16, 17] are targeted to the pre-synaptic and post-synaptic membranes, and FISSEQ is used to optically resolve and sequence the pre-synaptic and post-synaptic barcodes at a large fraction of synapses, thereby identifying connected pairs of cells in-situ. The idea of FISSEQ-BOINC is shown in **Figure 1**, and **Supplemental Note 7.1** describes possible strategies for targeting nucleic acid barcodes to the pre-synaptic and post-synaptic membranes.

The key challenges for FISSEQ-BOINC are fourfold:

**1) Biochemical cycling**: the large number of biochemical and imaging cycles required, e.g., at least 30 images to in situ sequence a 30-base cell-identifying barcode

**2) Resolution of distinct synapses**: the need to optically resolve a given synapse from nearby synapses, which will require sub-diffraction-limited optical microscopy and/or molecular stratification if *most or all* synaptic contacts are to be observed (whereas synapses from a sparse subset of neurons are routinely resolved with diffraction-limited optics)

**3) Resolution of pre-synaptic from post-synaptic barcodes**: the need to distinguish barcodes on the pre-synaptic and post-synaptic sides of a synapse, despite their close apposition across the synaptic cleft [18], which will require further targeted resolution enhancements

**4) Restriction of FISSEQ to synapses**: barcode RNAs localized in axons or dendrites would often co-inhabit resolution voxels with pre-synaptic and post-synaptic barcodes, implying a need either for precise targeting of barcodes *only* to synapses (and nuclei), or for restriction of the FISSEQ biochemistry itself only to synapses – optical resolution considerations [5, 11] further suggest that FISSEQ signal should be restricted as closely as possible to the synaptic cleft, rather than filling the entire pre-synaptic and post-synaptic compartments, i.e., the spine heads and axonal boutons

We next treat each of these challenges in turn.

**Figure 1. A fluorescent in-situ sequencing strategy for connectomics:** cell-identifying nucleic acid barcodes are targeted to the pre-synaptic and post-synaptic membranes, where their sequences are read by FISSEQ in a high-resolution optical microscope. Resolving synapses from their neighbors, and distinguishing pre-synaptic from post-synaptic barcodes at a given synapse, requires strategies for sub-diffraction optical imaging.

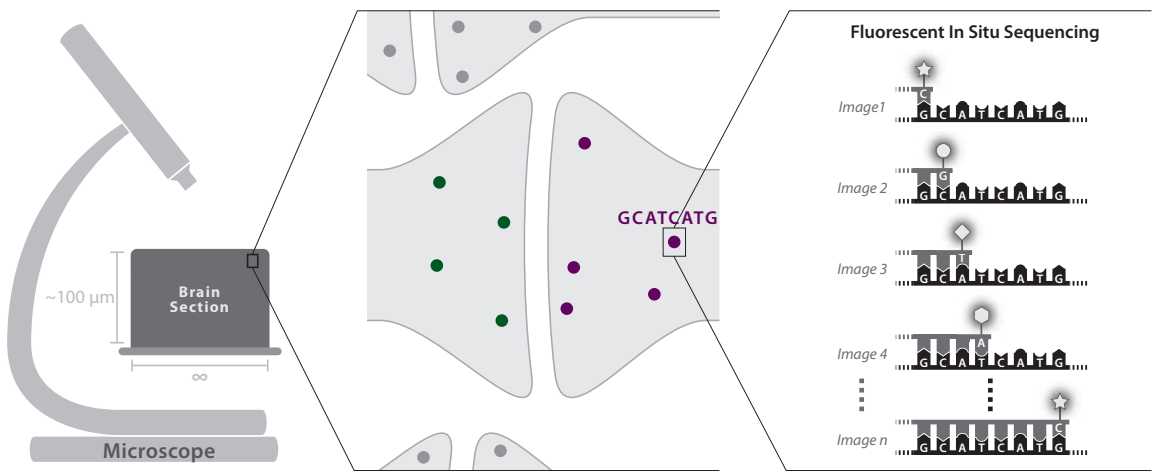

## 2.1 Biochemical Cycling

**Cycle time**  Current FISSEQ biochemistry steps – based on sequencing by ligation – take 2.5 hours per base, but using Illumina-type sequencing chemistries – based on sequencing by synthesis – this can be reduced to 30 minutes per base. A variety of alternative chemistries have been developed for fast cyclic sequencing by synthesis in polymerase colonies [19].

**Reagent cost**    Biochemical reagent costs are negligible compared to imaging costs (see below). Fluorescent ligation probes are the cost-limiting reagent in current FISSEQ protocols, available commercially for roughly $500 per $10^{17}$ molecules (and note that commercial biochemical reagents are often priced orders of magnitude above the synthesis cost). For comparison, we can estimate the number of ligation probe molecules required for whole-mouse-brain FISSEQ-BOINC as $n \cdot s \cdot m \cdot b \approx 10^{17}$ where $n \approx 10^8$ is the number of neurons, $s \approx 10^4$ is an upper bound on the average number of synapses per neuron, $m \approx 1000$ is the number of template copies per synapse, and $b \approx 100$ is the number of nucleotides per template. Thus, even if we require 100× excess probe molecules, the reagent cost per whole mouse brain synaptic FISSEQ-BOINC is less than $100k.

**Error rates**    Sequencing by ligation has typical error rates of roughly $\epsilon = 1\%$, such that a 30-base sequencing reaction has a success probability of $P_{\text{success}} = (1-\epsilon)^{30} = 74\%$. If each base is sequenced three times, however, a majority voting scheme can be implemented: the sequencing-associated error rate could thus be reduced to $3\epsilon^2 + \epsilon^3$, leading to $P_{\text{success}}^{\text{majority vote}} = (1-(3\epsilon^2+\epsilon^3))^{30} = 99.1\%$. This would triple the number of biochemical cycles. An alternative method to error-correct synaptic FISSEQ reads is discussed in **Supplemental Note 7.4**.

## 2.2 Resolution of distinct synapses

**Resolution requirement**    Diffraction-limited 3D optical microscopy ($\lambda/(2 \cdot \text{NA}) \approx 200\,\text{nm}$ $xy$-resolution and $2\lambda/\text{NA}^2 \approx 533\,\text{nm}$ $z$-resolution[2] for numerical aperture NA = 1.5 and wavelength $\lambda = 600\,\text{nm}$) is insufficient to reliably resolve nearby synapses, which are packed at an *average* density of 1-2 per $\mu\text{m}^3$ [5]. Prior theoretical studies [5, 11], constrained by EM anatomical data from rat hippocampal neuropil, suggest that $> 90\%$ of synapses could be resolved at $\sim 100\,\text{nm}$ isotropic resolution. This conclusion is subject to the assumptions used in the simulations, including fluorescent labeling only of the pre-synaptic and post-synaptic protein densities (PSDs) rather than of the entire synaptic compartment [5].

**Synapse-specific molecular labeling**    There exist multiple methods to optically label intact synapses via inter-cellular protein-protein interactions across the synaptic cleft, e.g., via neurexin-neuroligin interaction or split fluorescent protein complementation [22–24], or via immuno-staining against synapse-specific proteins [25, 26]. These methods could be used to validate the ability of an optical setup to resolve distinct synapses, and/or to locate synapses before in-situ sequencing. Because close axon/dendrite contacts do not reliably predict the locations of individual synapses [27], it would be desirable to use such an independent molecular marker of valid synapse locations, although we also invoke other methods here to eliminate FISSEQ signal that does not originate from actual synapses.

**Limits to spatial resolution from rolony size**    The rolling-circle nano-balls (rolonies) generated in FISSEQ are roughly 100–200 nm in diameter, in current protocols.

### 2.2.1 Strategies for resolution enhancement

We now consider strategies, which can be used alone or in combination, to ensure the resolution a given synapse from neighboring synapses: *super-resolution imaging*, *thin sectioning* and *molecular stratification*. These strategies are illustrated in **Figure 2**.

**Super-resolution imaging**    While much super-resolution microscopy research aims toward $< 10\,\text{nm}$ resolution and live-cell compatibility, FISSEQ-BOINC gives rise to a different set of challenges: 50–100 nm resolution in four colors, in fixed tissue, using standard fluorophores, and at the *highest possible speed*. The speed/resolution tradeoff is likely to be favorable in an in-situ sequencing context: rolonies are brighter than single fluorophores, and the protocol is robust to photobleaching because new dyes are flowed in on each cycle[3].

To perform FISSEQ, a microscopy platform must ideally allow 4-color imaging, or at least 3-color imaging. On a 3-color microscope, the 4th base could be unlabeled such that absence of signal serves as the fourth color. A 2-color microscope, however, is insufficient[4].

Among existing technologies, linear 3D structured illumination microscopy (SIM) [29] enhances resolution by a factor of 2 along all three axes, relative to the diffraction limit, and is naturally compatible with 4-color imaging using standard fluorophores. Analog SIM acquisition [30] can improve SIM speed and saturated SIM (SSIM) [31] can improve SIM resolution[5]. $I^5S$ two-objective detection [32] is a form of SIM with isotropic $\sim 100\,\text{nm}$ resolution. Other existing methods such as isoSTED can achieve $< 50\,\text{nm}$ resolution along all three dimensions [33], but may be more difficult to adapt to high-speed, 4-color operation.

---

[2]Note that methods like Bessel beam plane illumination [20] and diSPIM [21] give *isotropic* $\sim 300\,\text{nm}$ focusing.

[3]Oxygen-radical scavenging buffers [28] can also be applied to minimize photobleaching and photodamage.

[4]In theory, a scheme with 2× more fluorophore cleavage/removal cycles could be envisioned: flow on A and T, look with 2 colors, cleave and wash away the fluorophores without de-protecting the bases, flow on C and G, look with the same 2 colors, cleave and wash away the fluorophores, then de-protect and move to the next base. Unfortunately, in the current FISSEQ chemistry, fluorophore cleavage/removal is achieved via the same reaction as nucleotide de-protection. Use of a 2 color microscope would therefore require new FISSEQ chemistry.

[5]The light intensity of the sinusoidal illumination pattern needed for SSIM is >1 photon per cross-section per fluorescence lifetime; then the emission pattern is non-sinusoidal due to saturation and contains spatial information at higher harmonics.

**Figure 2. Resolution Enhancement Strategies for FISSEQ-BOINC:** super-resolution, molecular stratification, thin sectioning, and informatic deconvolution from a known barcode pool. These techniques can be applied alone or in combination to improve the resolvability of nearest-neighbor synapses, and/or of barcodes on opposing sides of the synaptic cleft. Super-resolution microscopes overcome the traditional diffraction limited resolution limit ($\lambda/2NA$) via a variety of methods, such as patterned illumination, nonlinear optical effects, or stochastic single-molecule blinking. Molecular stratification initiates FISSEQ of only a (random or pre-programmed) subset of molecular barcodes in each imaging frame, e.g., activating only pre-synaptic or only post-synaptic barcodes. Thin sectioning (physical or optical) allows enhanced lateral resolution in a 2D plane by eliminating overlaps in the third dimension. Informatic deconvolution decodes mixed FISSEQ signals from a combination of distinct barcodes within a single resolution voxel, by relying on prior knowledge of the pool of individual barcodes.

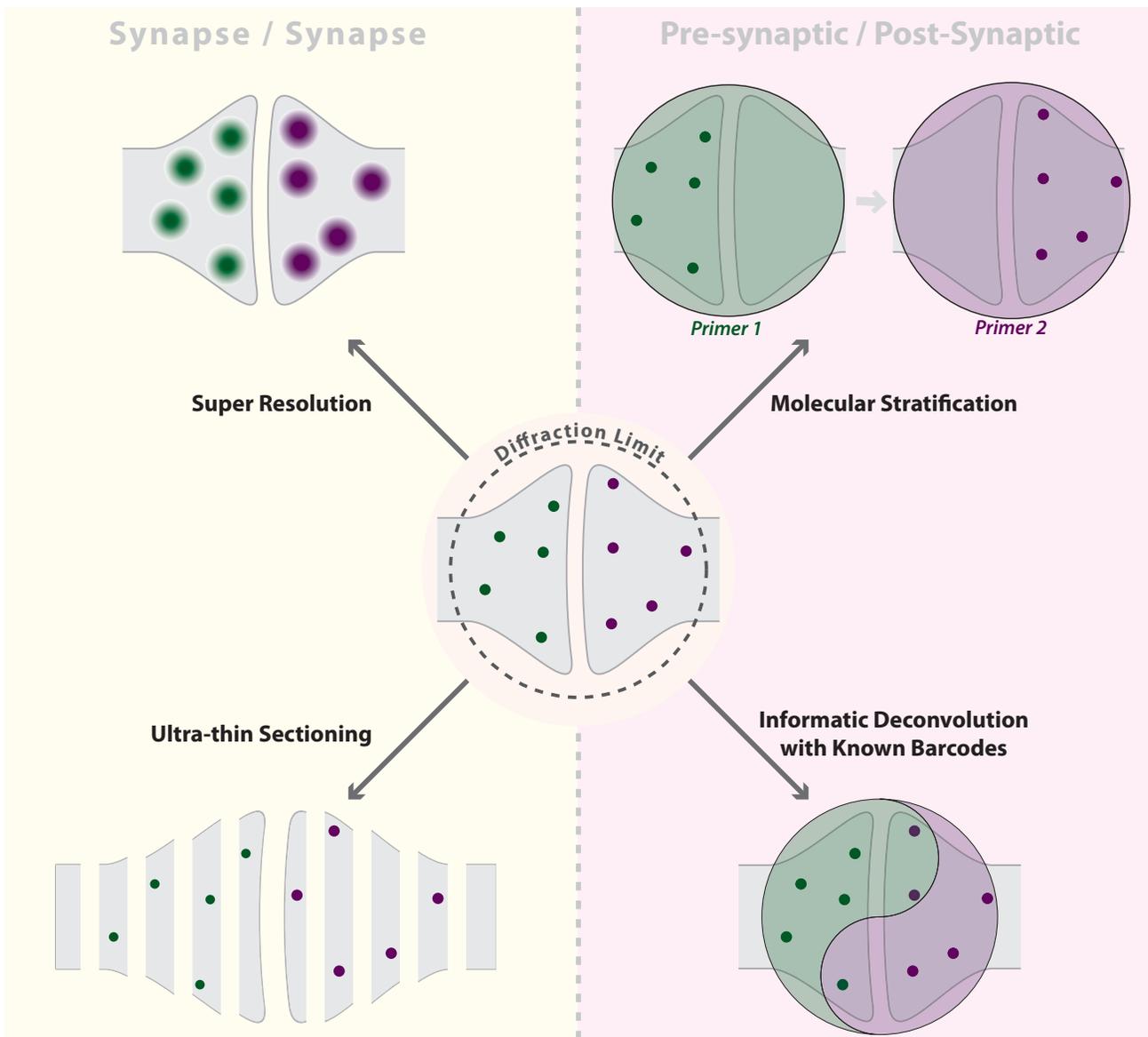

**Thin sectioning**     Thin sectioning below the intrinsic axial resolution of the imaging setup can improve the *effective* axial resolution, by physically separating otherwise unresolvable axial locations. It can also improve the effective lateral resolution, by decreasing the spot density in the $xy$ plane within each section. Experimentally, Array Tomography – a form of thin-sectioning microscopy – appears to resolve most synapses in mouse brain tissue [25, 26] using < 100 nm tissue sections on a diffraction-limited confocal microscope. In Array Tomography, the process of diamond-knife sectioning is automated, and an array of tissue slices is generated on a strip of adhesive for subsequent random-access automated imaging. All-optical methods can also achieve axial sectioning down to the 100 nm level [34, 35].

**Molecular stratification**     Molecular stratification is a method leading to linear improvements in the *effective* spatial resolution of FISSEQ with linear increases in the number of imaging cycles, at *fixed* optical resolution of the microscope, by activating only a *subset* of the molecules in each cycle [1]. In one implementation, multiple distinct "primer" sequences are employed sequentially during the FISSEQ process, such that each primer initiates sequencing of a subset of the rolonies.

There are two broad types of molecular stratification – stochastic and deterministic. In stochastic stratification, an *arbitrary* subset of barcodes is activated in each step; for example, all barcodes beginning with the nucleotide "C". In deterministic stratification, a *defined* subset of the barcodes is activated in each step; for example, all barcodes at pre-synaptic but not post-synaptic terminals, or all barcodes in L5 pyramidal cells but not in other types of neurons.

Stochastic stratification could be implemented via random ligation of primer binding sites to the target nucleic acids before FIS-SEQ [1]. Up to 4x stratification can be obtained just by varying the last base of the primer binding site, since polymerase initiation is strongly dependent on correct homology at this position. Further stratification can be achieved by working backwards, gaining additional stratification by a factor of 4 at each step. Stratification factors of 10-100 should be readily achievable with this approach, albeit with the corresponding increases in imaging time. Stochastic molecular stratification by a factor $S$ should lead to an increase in the average point-point separation by a factor of $S^{1/3}$ along all three axes.

Molecular stratification could be used to reduce the effective density of synapses in each imaging frame. To do so, however, all of the rolonies at a *given* synapse are activated or inactivated together. For example, stratification could be performed using the first few bases of the cell-identifying barcode itself, which is cell-specific rather than rolony-specific.

## 2.3   RESOLUTION OF PRE-SYNAPTIC FROM POST-SYNAPTIC BARCODES

To distinguish pre-synaptic from post-synaptic barcodes at a given synapse, further strategies must be employed for resolution enhancement of optical microscopy: barcodes positioned close to the pre-synaptic and post-synaptic membranes may be spaced apart by ~10–50 nm across the synaptic cleft [11, 18], well below the resolution of structured illumination microscopy and below the level where gains can be made through thin-sectioning. Molecular stratification and *informatic deconvolution* can solve this problem, as depicted schematically on the right side of **Figure 2**.

**Molecular stratification**     In an elegant implementation of deterministic stratification, pre-synaptic and post-synaptic barcodes could be fused to distinct primer binding sites. A first primer would drive in situ sequencing only of pre-synaptic barcodes; then, in a subsequent set of biochemical cycles, a second primer would drive in situ sequencing only of post-synaptic barcodes. This would *eliminate the need to resolve pre-synaptic from post-synaptic barcodes in any single fluorescent image*. **Supplemental Note 7.2** details a range of possible strategies to implement deterministic molecular stratification at the synapse.

To summarize: if a given synapse is already resolvable from *other* nearby synapses, then additional deterministic molecular stratification by a factor of 2 – by using distinct pre-synaptic and post-synaptic priming sequences – would be sufficient to resolve pre-synaptic from post-synaptic barcodes.

**Informatic deconvolution**     So far, we have assumed that the pre-synaptic and post-synaptic barcodes must be optically resolved, either in the same imaging frame, or in separate imaging frames (via molecular stratification). This is not necessary, however, if the set of *possible* pre-synaptic and post-synaptic barcodes is already known, and is a small subset of the $4^n$ DNA strings of length $n$, where $n$ is the barcode length.

For example, suppose that each cell's barcode is readable via FISSEQ in the cell's nucleus. In a first step, one can then perform FISSEQ on all the cell nuclei. One thereby determines all possible pre-synaptic or post-synaptic barcode sequences. Now suppose that both pre-synaptic and post-synaptic barcodes are contained within a single resolution voxel at each synapse, such that "mixed" fluorescent signals are obtained from the synapse during FISSEQ. Given the mixed signal, and the known set of possible barcode sequences obtained from nuclear FISSEQ, it is then possible to determine computationally which *combination* of the known barcode sequences gave rise to the observed mixed signal[6].

---

[6]Additionally, synapse *locations* could be pre-labeled with an antibody (e.g., Anti-Synapsin I) and optically resolved, so that the microscope knows "where to look" for such mixed signals.

We refer to this approach as *informatic deconvolution* of non-resolvable barcode combinations, given a pre-determined pool of individual barcode sequences. **Supplemental Note 7.4** outlines options for informatic deconvolution in more detail.

## 2.4 Restriction of FISSEQ to synapses

Restricting the FISSEQ signal to synapses is critical to the implementation of FISSEQ-BOINC. Barcodes localized "part-way down" axons or dendrites would often co-inhabit FISSEQ resolution voxels occupied by genuine synapses. Furthermore, to achieve sufficient resolvability of neighboring synapses, the FISSEQ signal should be restricted as closely as possible to the synaptic cleft [5, 11].

Methods to accomplish this restriction fall into two categories: barcode localization (trafficking RNA barcodes solely to the synapse) and location-restricted sequencing (restricting FISSEQ chemistry solely to the synapse). **Supplemental Note 7.3** describes one strategy for restricting FISSEQ biochemistry to the synapse, even in the presence of barcode RNAs localized outside the synapse.

## 2.5 Cost estimate: Ultra-thin sectioning FISSEQ-BOINC

One approach would be to perform FISSEQ on < 100 nm thin tissue sections in a standard diffraction-limited microscope, such as a confocal microscope. We treat this case here merely because it leads to a simple calculation of the estimated imaging cost; other methods allowing comparable degrees of super-resolution, which do not rely on thin sectioning, may be preferable in practice.

**Spatial resolution**    Experimentally, this approach appears to resolve most synapses in the context of Array Tomography [25, 26]. Resolution calculations [11] indicate that Array Tomography at 100 nm vertical slice thickness and 200 nm lateral resolution (via a standard confocal microscope) would resolve >90% of synapses from their nearest neighbors.

**Imaging time**    A comparison with the rates of sequencing by fluorescent microscopy in commercial sequencing machines suggests that FISSEQ-BOINC in < 500 nm thin sections could proceed at rates of 8 hours per cm$^2$ [5]. Assuming 100 nm sections, the total imaging time is

$$T_{mouse}^{ultra-thin\ sectioning\ FISSEQ} = \frac{8\ hours}{100\ nm\ section} \times 42000\ sections = 40\ years$$

to sequence >90% of mouse brain synapses at Illumina rates[7] on a single machine [5].

**Imaging cost**    With $1M in situ sequencing machines operating at Illumina speeds, and machine cost amortized over three years, the imaging cost would be $13M per whole mouse brain connectome[8].

**Additional costs**    The cost of instrumentation for whole mouse brain sectioning and section-handling, at 100 nm slice thickness, can likely be reduced below $1M [5].

**Project cost**    The cost for a 3-year ultra-thin sectioning mouse brain FISSEQ-BOINC is therefore in the $10M-$20M range, subject to the prior experimental demonstration of the basic molecular mechanisms of FISSEQ-BOINC. This cost could be reduced if the individual microscopes were brought below $1M, or amortized over multiple projects.

**Optical vs. physical sectioning**    Via all-optical techniques, $I^5M$ microscopy achieves 100 nm axial and roughly 200 nm lateral resolution in a wide-field mode [34]. Parallelized multi-photon 4Pi-confocal microscopy leads to similar axial resolution [35]. While we have focused on the physical sectioning approach here, because its ability to resolve densely labeled synapses has been demonstrated experimentally [25, 26] and because it leads to a simple estimate of the imaging cost, optical sectioning may be preferable in practice, e.g., if auto-alignment methods for 4Pi interference are introduced.

## 2.6 Hybrid approaches

Hybrids between the bulk sequencing (BOINC) [3–5] and in situ sequencing (FISSEQ-BOINC) could lead to reduction of total costs, as well as potential experimental simplifications. For example, FISSEQ could be performed solely on each cell's self-barcode – localized in the nucleus – and BOINC bulk-sequencing could subsequently be used to determine the connectivity matrix. This would eliminate the requirement for high resolution direct observation of synapses by light microscopy, yet would still allow localization of each cell body in addition to determination of the connectivity matrix. **Supplemental Note 7.5** discusses a potential hybrid strategy.

---

[7]For comparison, consider a hypothetical high-speed super-resolution imaging system operating at 50 nm isotropic resolution and 1 giga-pixel per second frame rate. The time for 30-base in-situ sequencing of whole mouse brain is $30 \times 2^3 \times 420\ mm^3/(50\ nm)^3 \times 1\ s/10^9 = 26$ imaging years, where the factor of $2^3$ is for Nyquist sampling in 3D.

[8]To distinguish pre-synaptic from post-synaptic barcodes, either a deterministic molecular stratification approach would be used – with distinct pre-synaptic and post-synaptic primer sequences, increasing the imaging cost by a factor of 2 – or informatic deconvolution could be applied based on a known barcode pool, such as from nuclear FISSEQ, without any increase in the imaging cost.

## 2.7 Summary

FISSEQ-BOINC leverages the recent development of fluorescent sequencing protocols for nucleic acids that have been locally amplified inside intact tissue [1, 2]. This approach could have several potential advantages over other connectomics approaches such as axon tracing via large-scale serial-section electron microscopy. Most notably, like BOINC [3], it does not require error-prone morphological tracing of thin (< 100 nm) neural processes over large (~1 cm) distances, instead relying on a digital representation of cell identity in nucleic acid strings for which *the fidelity of readout is independent of distance from the neuronal soma*.

FISSEQ-BOINC could allow the direct observation of synaptic connections in situ by reading the sequences of pre-synaptically and post-synaptically localized cell-identifying nucleic acid barcodes. Achieving the necessary spatial resolution in fluorescent microscopy, however, requires a suitable combination of: a) super-resolution microscopy, b) physical [25, 26] or optical [34] sectioning of tissue into ~100 nm slices, c) molecular stratification, and/or d) informatic de-convolution of multiple sequences within one optical resolution voxel using a known barcode pool.

A preliminary cost estimate suggests that FISSEQ-BOINC in 100 nm ultra-thin 2D sections may be achievable for roughly $10M per three year mouse connectome, using imaging equipment comparable to today's bulk fluorescent sequencing machines. The ultra-thin sectioning approach has been demonstrated experimentally to resolve densely labeled synapses in mouse cortex, and may integrate naturally with Array Tomography immuno-staining methods that report on the molecular diversity of synapses [25, 26].

FISSEQ-BOINC integrates readily with other light-microscopy-based readouts, conferring high-dimensional, molecularly specific information: FISSEQ-BOINC is thus naturally suited to the acquisition of Rosetta Brain datasets, as described below. In the simplest case, this includes integration with other FISSEQ readouts from the same tissue specimen, such as in situ transcriptomics [36] or in situ readout of cell lineage barcodes.

In the specific implementation proposed here, which sequences only synapse-localized and nuclear-localized barcodes – FISSEQ-BOINC does not trace the morphologies of neurons. It is possible that whole-cell FISSEQ could recover detailed morphology, however, if each neuron is filled sufficiently with barcoded transcript, in a manner somewhat analogous to current BrainBow approaches [13–15].

# 3  Preliminary experimental directions for FISSEQ-BOINC connectomics

## 3.1 Technology components

Given the prior experimental demonstration of FISSEQ [1, 2], key experimental milestones on the way towards FISSEQ-BOINC include the following.

- Barcode each neuron with a unique RNA tag. For dense, whole-brain barcoding, the barcode generation mechanism should be genomically encoded rather than delivered virally. Mouse models may be readily accessible due to their ease of genetic manipulation, e.g., via embryonic stem cell implantation. A germline-competent transgenic encoding the barcoding mechanism must be developed, despite the fact that expressed barcodes pose potential issues of toxicity. Due to its small brain size, with only a few million cortical neurons [37], the Etruscan shrew may also be a desirable target.

- Find a set of RNA localization tags that can label all synapses regardless of cell type. This is discussed in **Supplemental Note 7.1**.

- Restrict the FISSEQ signal to the pre-synaptic and post-synaptic densities (PSDs): a FISSEQ enzyme such as phi29 could be targeted to the PSDs via an antibody (localization to entire synaptic compartments is not sufficient [5] to allow good resolvability of neighboring synapses in a dense labeling scenario). This is discussed in **Supplemental Note 7.3**.

- Demonstrate optical resolution enhancement in the context of FISSEQ: to observe *nearly 100% of synaptic contacts* unambiguously, a 4-color (or at least 3-color) super-resolution microscopy with 50–100 nm $xy$ resolution is needed, coupled with < 100 nm thin sectioning or < 100 nm axial resolution. Molecular stratification is a complementary tool for optical resolution-independent enhancement of the *effective* spatial resolution, and methods for molecular stratification are discussed in **Supplemental Note 7.2**.

- Achieve chemical compatibility of FISSEQ with appropriate tissue-embedding and immuno-staining reagents: compatibility of FISSEQ with Array Tomography preparations would be valuable (**OPTIONAL**).

- Achieve biochemical compatibility of bulk BOINC with FISSEQ to allow hybrid strategies: BOINC could be used to obtain connectivity information via bulk sequencing while FISSEQ could be used on the same sample to obtain cellular positions and transcriptomes (**OPTIONAL**).

## 3.2 FISSEQ-BOINC projectomes

Simplified versions of FISSEQ-BOINC could be useful even while relaxing some of the above engineering requirements. For example, rather than mapping the precise synaptic connectivity, a first implementation could target FISSEQ-BOINC barcodes only to the pre-synaptic compartments and nuclei, thus obtaining the complete *projectome* of a single brain, i.e., the spatial locations to which all axons project. This would obviate the need to resolve pre-synaptic from post-synaptic compartments, requiring only FISSEQ of pre-synaptic compartments and donor nuclei / somas. Projectomes are powerful resources [38–40] of interest to many neuroscientists, and are useful for constraining theories of brain architecture [41], yet current approaches require integrating many experiments across many brains and do not reach single-cell precision [42]. FISSEQ-BOINC would solve both problems.

Note that cellular-resolution, single-brain projectomes could also be achieved with BOINC through 3D sectioning of the tissue and subsequent section-specific barcoding; nuclear-localized DNA barcodes (revealing soma positions) could be distinguished from synapse-localized RNA copies of the barcodes (revealing projections) by observing the removal of introns or by other methods.

# 4 Toward Rosetta Brains

A key goal for neuroscience is to measure many biological variables simultaneously, in a co-registered fashion, within *single* brains [6]. Even co-registering just *two* variables at cellular resolution (e.g., activity and connectivity, or gene expression and projection pattern) has led to insights inaccessible to separate measurements [40, 43], and it is important to extend such co-registration to as many relevant variables as possible. The results of such rich co-registration would constitute *Rosetta Brains*: integrative data-sets that could constrain theoretical efforts to bridge across levels of structure and function in neural tissue.

Such integration could be enabled by the emerging ability to translate a variable of interest, such as synaptic connectivity, cell lineage [44], or perhaps even dynamic activity patterns [45–47], into a physical form that can be robustly stored, transported and measured – digitally-encoded nucleic acid strings, which can be read out in situ via FISSEQ as temporal patterns of colors.

## 4.1 FISSEQ-BOINC approach to Rosetta Brains

We can envision at least one route to Rosetta Brains, as follows [6]. Given a single brain, we would like to measure "A, B, C, D and E": **A**ctivity, **B**ehavior, **C**onnectome, **D**evelopment and gene **E**xpression. All of these could potentially be obtained in the context of a FISSEQ-BOINC approach, as follows:

**A**ctivity can be measured via electrodes, optical microscopy, molecular recording [45–47] or other techniques [48, 49], in a manner that allows co-registration [43] at the single-cell level with subsequent fixed-tissue optical microscopy.

**B**ehavior can be monitored via video in the context of hypothesis-driven experimental paradigms or free behavior.

**C**onnectomes can be obtained via FISSEQ of nucleic acid cell ID's indicating connectivity, "ID-C", as well as FISSEQ of nucleic acid barcode-tagged antibodies targeted to specific synaptic proteins, revealing synapse properties [50].

**D**evelopmental lineage [44, 51] can be determined via FISSEQ of nucleic acid cell ID's encoding cell lineage:"ID-D", e.g., DNA barcodes which are "updated" once per cell cycle [52, 53].

**E**xpression of genes is measured in a spatially-resolved fashion via FISSEQ or fluorescent in-situ hybridization [54, 55] of mRNA and/or of nucleic acid barcoded antibodies targeted against cellular proteins: "ID-E".

Implementing such a strategy would entail performing behavioral experiments on a single animal, with measurement of activity data occurring in real time. Then, the brain tissue would be fixed and thin-sectioned [25, 26, 56], perhaps using methods similar to those of Array Tomography [26]. The tissue sections would be subjected to highly multiplexed FISSEQ and immuno-staining cycles in a high-speed, high-resolution optical microscope.

## 4.2 Advantages

**Natural multiplexing of immuno-labels and transcripts**     Attaching each protein-probing antibody to a unique nucleic acid barcode allows multiplex in-situ readout of synaptic and cellular proteins (exponentially parallelized compared to the typical 4-8 fluorescent colors per staining cycle).

Furthermore, in-situ sequencing of mRNA is inherently multiplexed. It will likely be possible with FISSEQ to achieve orders of magnitude higher multiplexing compared with recent gene expression atlases (8 hybridization probes per brain in 25 $\mu$m sections) [57], while performing these analyses on single brains rather than populations of brains.

**Error-correction**   By combining many types of measurements on a single piece of tissue, there is the possibility for error-correction: for example, mRNA expression levels could cross-validate cell type inferences made on the basis of synaptic protein abundances and distance-dependent connectivity patterns.

Furthermore, an array of molecular barcoding techniques could cross-validate or even substitute for in-situ sequencing, e.g., multiplexed in-situ probe hybridization [54, 55][9].

**Combining structural and dynamic measurements**   During thin sectioning, electrodes or other recording devices [48] could potentially remain inside the tissue, with the microtome simply slicing through them – although damage to the knife would be a serious concern, perhaps necessitating soft, thin electrodes. The positions of the electrodes or other recording devices could be known post-facto from microscopy or pre-facto from high-resolution CT scans. This would allow molecular composition and connectivity to be ascertained for cells of known activity history and representation/coding properties. Optical imaging of neural activity would integrate even more readily with fixed-tissue optical microscopy for FISSEQ. Much more speculatively, molecular recordings of time-dependent signals [45, 46] could be read out directly though FISSEQ or other forms of optical microscopy [47].

## 4.3   Limitations and extensions

Measuring morphology is important, since mechanisms such as ephaptic coupling [60], BDNF signaling [61], nitric oxide release [62] and many others can functionally link nearby neurons in a distance-dependent manner. Also, the positions of inputs within the dendritic tree govern their electrical integration [63, 64]. Therefore, knowledge of the precise spatial appositions of neuronal processes seems relevant for understanding elementary neuron-neuron interactions. Variants of FISSEQ-BOINC, employing FISSEQ barcodes distributed *throughout* all cellular compartments, could reveal at least the gross morphology of every cell, much as is done in current BrainBow techniques [13–15]. Certainly, a genetically targeted subset of cell shapes could be imaged in this way. In addition, correlative [65, 66] light microscopy (for FISSEQ) and EM (for high-resolution morphology and ground-truth connectivity) could be used for validation, if the tissue preparation chemistries associated with EM and FISSEQ can be made compatible (as has been done for Array Tomography and SEM [25]).

Neuromodulators (e.g., neuropeptides and biogenic amines) can strongly modify the behavior of neural circuits, effectively forming "circuits within circuits" that are activated or inactivated by various modulators [67, 68]. Because many neuromodulatory receptors [69] and release sites [70] are extra-synaptic [67, 71], mapping the neuromodulatory circuitry will require more than just *synaptically-localized* immuno-staining. While this goes beyond the pure synaptic FISSEQ-BOINC approach described above, this could also be done using optical imaging and multiplexed antibody staining in the same setup. Alternatively, correlations between gene expression patterns and neuromodulatory responses could be measured independently, and then used to *infer* single-cell neuromodulatory properties from FISSEQ transcriptomics data.

We have not discussed how to measure the *dynamics* of connectomes, or of the associated molecular annotations. This will require additional new concepts, perhaps variants of existing molecular recording ideas [45, 46, 48], but tailored to particular (e.g., slower) timescales and processes of interest (e.g., transcription). Integrating activity data into Rosetta measurements raises highly non-trivial technology problems if it is to be done at scale [48].

# 5   Discussion

FISSEQ-BOINC is a hypothetical connectomics strategy which uses Fluorescent In-Situ Sequencing (FISSEQ) to directly read the sequences of co-localized RNA barcodes at the synaptic cleft. Once the basic molecular mechanisms are established, preliminary estimates indicate that a FISSEQ-BOINC analysis of an entire mouse brain could cost $10M-$20M for a three-year project, primarily in the form of microscopy equipment [5]. Improvements to the speed of 50–100 nm 4-color super-resolution fluorescence microscopy in fixed-tissue could reduce this cost further.

FISSEQ-BOINC is only one example of a plausible light microscopic strategy for high-speed, molecularly annotated connectomics. Unforeseen limitations of this plan could be revealed experimentally and alternative designs may prove superior. Nevertheless, the FISSEQ-BOINC strategy illustrates the existence of a flexible emerging design space. Ultimately, such an approach could simultaneously provide spatial localization of neuronal somas, determination of the neuron-neuron connectivity matrix, and cellular-resolution molecular annotations indicative of cell types, synapse properties, and developmental lineages. Furthermore, molecularly annotated connectomics approaches could synergize with strategies – such as the Human Brain Project [72, 73] – that aim to build large-scale data-driven simulations via integration of diverse measurements and data-sets.

---

[9]Fluorescent in-situ hybridization approaches have demonstrated experimentally a 32-fold simultaneous multiplexing capacity using Binomial(7,3) = 35 STORM activator/emitter photo-switchable pairs [58]. This approach could theoretically scale to at least 792 effective colors using available emitters, or to tens of thousands of effective colors if an infrared (IR) fluorophore was added to the color palette or other methods were used to distinguish dye pairs [59].

Several recent and ongoing developments converge to enable the possibility of FISSEQ-BOINC or similar strategies. The problem of tracing thin axons over large distances though vast image stacks is, in principle, obviated by the use of digitally-encoded biopolymers which can be *physically* transported along the axon by endogenous cellular mechanisms. This combinatorial sequence space can be accessed by using $N$ cycles of fluorescence-coupled biochemistry to read out $4^N$ effective "colors". Advances in light microscopy beyond the diffraction limit – which could be as simple as $\sim 100$ nm physical or optical sectioning – allow spatial resolution of many of the key objects of interest within any single fluorescent image, but further improvements in the *effective* spatial resolution can be obtained by stratifying otherwise unresolvable objects into *separate* image frames using flexible in-situ DNA manipulation techniques.

FISSEQ-BOINC is currently at the level of a theoretical proposal. The implementation of FISSEQ-BOINC or related strategies will require the solution of a number of experimental challenges in in-situ biochemistry, imaging and automation. None of these appears to be insurmountable, however, and ongoing advances in fields such as high-speed super-resolution microscopy and automated tissue sectioning could remove some of the existing technical obstacles. We can thus envision at least one path, which may be one among many viable alternatives, towards whole-brain-scale, molecularly annotated connectomics.

## 6 ACKNOWLEDGMENTS

We thank Paul Goodwin from GE / DeltaVision for discussions on optical microscopy, Robert Barish and Ralf Jungmann for discussions on single-molecule super-resolution techniques, and William Shih and Peng Yin for discussions on nanoscale barcoding. We thank Fei Chen, Paul Tillberg, Ian Choi, Gary Marcus, Dario Amodei, Ted Cybulski, David Dalrymple, Tom Dean, Noah Donoghue, Kevin Esvelt and Brad Zamft for discussions. We also thank David Dalrymple for the use of his LaTeX template.

Adam Marblestone is supported by the Fannie and John Hertz Foundation fellowship. Konrad Kording is funded in part by the Chicago Biomedical Consortium with support from the Searle Funds at The Chicago Community Trust. Ed Boyden is supported by the National Institutes of Health (NIH), the National Science Foundation, the MIT McGovern Institute and Media Lab, the New York Stem Cell Foundation Robertson Investigator Award, the Human Frontiers Science Program, and the Paul Allen Distinguished Investigator in Neuroscience Award. Reza Kalhor, Evan Daugharthy, Seth Shipman and George Church acknowledge support from the Office of Naval Research and the NIH Centers of Excellence in Genomic Science. Seth Shipman is supported by the National Institute on Aging (5T32AG000222-22). Yuriy Mishchenko acknowledges support from Bilim Akademisi, the Science Academy, Turkey, under the BAGEP program, and support from BAP Scientific Research Projects Fund of Toros University. Tony Zador, Ian Peikon and Justus Kebschull acknowledge an NIH TR01 award and the Paul Allen Distinguished Investigator award. Justus Kebschull acknowledges a PhD fellowship from the Boehringer Ingelheim Fonds.

# 7  SUPPLEMENTARY INFORMATION

## 7.1  IMPLEMENTING BARCODE-TRAFFICKING TO THE SYNAPSE

While each of the below approaches will likely lead to an enhancement of synapse-localized barcodes, compared to passive diffusion, none is likely to yield the stringent synapse-specificity required for FISSEQ-BOINC: some fraction of barcode RNA will remain localized *outside* the synapse. Thus, one of these approaches will need to be combined with a secondary spatial restriction such as location-restricted FISSEQ, as outlined in **Section 7.3**.

**Barcode targeting via protein fusions**    A cell-barcode in the genome could be transcribed into RNA molecules bearing either the MS2-binding RNA aptamer sequence or the PP7-binding RNA aptamer sequence (via direct genetic fusion or RNA-RNA interaction [74]). The cytoplasmic domain of the pre-synaptic protein neurexin could be fused to MS2 protein, and the cytoplasmic domain of the post-synaptic protein neuroligin could be fused to PP7. These protein-RNA interactions are quite strong, e.g., with a 3 nM affinity constant for the MS2 aptamer/MS2 protein interaction [75]. Alternatively, RNA barcodes could perhaps be fused covalently to the targeting proteins via cap-snatching protein mutants from viruses [76] (a form of in-vivo mRNA display) or engineered ribozymes [77]. The barcode RNAs could thereby be localized to the pre-synaptic and post-synaptic compartments.

The difficulty for this approach lies in the molecular diversity of synapses. Taking the neurexin/neuroligin example, there are four neuroligin genes in the mouse. The different subtypes of neuroligin can be found at different subsets of synapses, with neuroligin 1 at excitatory synapses and neuroligin 2 at inhibitory synapses. Neuroligin 3 is found at both excitatory and inhibitory synapses, but likely not at every synapse in the mouse brain. The C-terminal tails of the various neurexins/neuroligins are likely sufficient to direct targeting of the appropriate fusion proteins, although the cell-type context dependence of this targeting would need to be tested.

Thus, this approach may require not just two MS2/PP7-tagged synaptic proteins such as neurexin and neuroligin (i.e., one pre-synaptic and one post-synaptic), but rather a small set of MS2/PP7-tagged pre-synaptic proteins and small set of MS2/PP7-tagged post-synaptic proteins, to cover the molecular diversity of all synapses. There are likely other groups of (e.g., non-membrane) proteins which *collectively* could be used to target barcodes to *all* synapses: candidates include PSD95, gephryin, SHANK3, Synapsin and others.

In this scheme, the cell barcode itself need only bear the MS2 or PP7 aptamer sequence; a transgenic mouse could be created (e.g., with CRISPR) that fuses the MS2 or PP7 proteins to multiple endogenous pre-synaptic or post-synaptic proteins.

**Barcode targeting via FINGRs**    Another targeting strategy would utilize fibronectin intrabodies generated with mRNA display (FINGRs) [78]. These are a form of genetically-encoded "protein aptamer" against a target protein of interest, whose expression level is additionally tuned to be at or below the level of the target, leading to highly specific labeling. FINGRs have already been generated against Gephyrin and PSD95 [78].

**Barcode targeting via endogenous dendritic or axonal mRNA localization signals**    Sequence signals in the 3'-UTR of mRNAs direct their localization to neuronal processes, e.g., the dendrites, via the formation of ribo-nucleoprotein "granules" which are transported by cytoskeletal motor proteins to their appropriate sub-cellular destinations [79]. Further sequence signals in the 5'-UTR appear to be sufficient to direct localization to the synapse [80], in some scenarios, although the full molecular underpinnings of subcellular RNA localization are not well known. Synapse associated poly-ribosome complexes, which perform spatially-localized protein translation at synapses, appear to be tightly localized to the synapse, and in particular to the base of the dendritic spine [81].

In principle, the appropriate RNA localization tags could be appended to cell-barcode RNAs to direct their localization to the pre-synaptic or post-synaptic densities. If multiple distinct localization tags must be appended, in a mutually exclusive fashion, to a single stochastically-generated RNA barcode, RNA trans-splicing or transposon traps could be used, as discussed in **Section 7.2**. Unfortunately, very few RNAs appear to localize exclusively to the synapse itself [82] let alone to the PSDs, so this will necessitate combination with the methods of **Section 7.3** or similar.

## 7.2  IMPLEMENTING MOLECULAR STRATIFICATION AT THE SYNAPSE

**Implementing molecular stratification via alternative RNA splicing**    Individual cells can *simultaneously* express at least 14-50 RNA splice variants of the same gene [83]. For deterministic molecular stratification to distinguish pre-synaptic from post-synaptic barcodes, two mutually-exclusive exons could be placed downstream of the barcode RNA: for example, one exon could contain the MS2 aptamer and reverse transcription priming site 1, while the other exon could contain the PP7 aptamer and reverse transcription priming site 2. Then, the same (e.g., genomically-encoded) DNA barcode sequence would give rise to two distinct RNA sequences, one targeted to the pre-synaptic compartment and specific to the first primer, and the other targeted to the post-synaptic compartment and specific to the second primer (using different sequencing adaptors might be preferable to different reverse-transcriptase primer binding sites, so that rolony preparation could occur in a single step).

In another implementation of the same idea, the cell could express multiple alternative RNAs which could be trans-spliced (rather than cis-spliced) onto a given barcode RNA. This has the advantage that alternative exons do not need to be engineered into a single gene [84]. The efficiency of trans-splicing would likely be a limiting factor here, especially since un-spliced targeting RNAs could saturate the available synaptic proteins.

**Implementing deterministic molecular stratification via site-directed gene duplication** Alternatively, gene duplication events could be targeted to a predetermined genomic location, e.g., via a "transposon trap" [85]. A barcode would first be generated within the retrotransposon sequence at a defined "start-site" location in the genome. Elsewhere in the genome would be placed two "landing sites" of the form: Promoter - Landing Site - PreTag/PostTag, where PreTag and PostTag are targeting sequences for pre-synaptic or post-synaptic localization. The retrotransposon can then be induced to "copy-and-paste" the barcode to these predefined landing sites. The landing sites would be known sequences where that transposon has a high probability of landing. Transposon-based duplication might have the advantage of being less dependent on cell-type-specific regulation as compared with alternative splicing.

**Implementing molecular stratification via overlapping genes (plus vs. minus strands)** Alternatively, to implement deterministic two-fold molecular stratification, the same DNA encoded cell barcode could be transcribed from one promoter in the forward direction – leading to one set of primer binding sites and localization tags – and from another promoter in the reverse direction, appending the second set of primer binding sites and localization tags. For a sufficiently diverse barcode space, the reverse complement of a given barcode in the pool would not itself be present in the pool, meaning that a cell could be identified *either* by its forward or reverse-complemented barcodes. It might be necessary to prevent spurious hybridization of the forward and reverse-complemented barcode RNAs (leading to spurious generation of RNA barcode dimers as well as PKR-activating long dsRNAs [86]), but it should be noted that many opposite-strand overlapping genes occur naturally in the human and mouse genomes [87].

## 7.3 Restriction of FISSEQ biochemistry to the synapse

Additional synapse specificity of the FISSEQ signal could be achieved by restricting the sequencing chemistry to the pre-synaptic and post-synaptic densities using a form of protein tagging. This could be achieved by conjugating the various enzymes used in the FISSEQ rolony preparation to antibodies against synaptic proteins. Such enzyme-antibody conjugates are not unreasonable given the success of previous enzyme antibody conjugates such as horseradish peroxidase, urease, and alkaline phosphatase. Moreover, by multiple conjugating enzymes in successive steps of the FISSEQ process to antibodies targeting different synaptic proteins found at the same synapse (e.g. MMuLV RT — anti-PSD95, CircLigase — anti-NR1, Phi29 — anti-NLGN3), spurious reactions at non-synaptic sites could be progressively reduced. Enzyme-antibody conjugates could be applied at a non-permissive temperature (or in the absence of substrate), unbound copies washed away, and then switched to a permissive temperature for the reaction.

Alternatively, a genetically-encoded non-specific biotin ligase [88], fused to a synaptic protein, could be used to biotinylate the proteins at the synapse, in a strategy similar to that of [89]. One of the FISSEQ enzymes could then be linked to streptavidin. After enzyme addition and a wash step to remove unbound enzyme, enzyme activity would be restricted to the synapse, such that FISSEQ signals would only originate from the synapse itself.

Note that protein localization at the synapse, as is exploited in this strategy, can be highly precise, perhaps due in part to synapse-localized translation in some cases [81, 90].

## 7.4 Prospects for informatic deconvolution

In the case where two DNA molecules are sequenced within a single optical resolution voxel (e.g., a diffraction-limited spot), the information content of each base sequenced is 3.25 bits[10] (whereas 4 bits would be recovered if it could be known which amplicon generated which signal). *Informatic deconvolution* is the process of providing additional information in order to make up for the missing 0.75 bits and generate a full sequence pair. This additional information can come from two sources: priori knowledge of all sequences in the pool, or additional "programmatic sequencing" reactions. It would also be necessary to append additional sequence tags to the barcodes, indicating which barcodes are pre-synaptic and which are post-synaptic. Otherwise, only an unordered pair of sequences is attributed to each synapse.

---

[10]Four of 16 possibilities (AA, TT, CC or GG) each convey 4 bits of information since the bases of both strands are disambiguated, whereas the remaining 12 possibilities convey only $4 - 1 = 3$ bits of information since the bit specifying which base goes with which strand is left ambiguous. Thus the Shannon entropy is on average $\frac{4}{16} \cdot 4 + \frac{12}{16} \cdot 3 = 3.25$ bits per base sequenced.

**Known barcode pool**     Consider first the problem of adding information from the known barcode pool. Let $n$ be the number of neurons (unique, known barcodes), $s$ be the number of synapses observed and $N$ be the barcode length. In a given resolution voxel (e.g., synapse) containing both pre-synaptic and post-synaptic barcodes, the observation consists of a list of unordered pairs corresponding to the labels at each base, such as "(A and T) or (T and A)".

Denote the observation as $x$. There are $< 2^N$ sequence pairs consistent with $x$, given no prior knowledge: the scaling will be $(\frac{1}{4} \times 1 + \frac{3}{4} \times 2)^N = 1.75^N$ on average. These are embedded in a space of $4^{2N}$ possible sequence pairs independent of the observation. We know that at least one pair of barcodes, each chosen from the set of size $n$ of known barcodes, is consistent with $x$. Perfect deconvolution is possible if and only if there is no other such pair which is also consistent with $x$.

For random sequences, the probability that any other pair of known barcodes is also, by chance, consistent with $x$ is

$$< \frac{n^2 \cdot 2^N}{4^{2N}} = n^2 \cdot 2^{-3N}$$

Since there are $s$ synapses in the brain, the expected number of ambiguous synapses is then $< s \cdot n^2 \cdot 2^{-3N}$.

For a mouse brain with $s < 10^{12}$ and $n = 10^8$, the expected number of ambiguous synapses via informatic deconvolution of pre-synaptic and post-synaptic barcodes is at most $S_{\text{ambiguous}} = 10^{28} \cdot 2^{-3N}$. For $N = 30$, $S_{\text{ambiguous}} = 8$ whereas $S_{\text{ambiguous}} = 0.0002$ for $N = 35$. Therefore, 30-35 base barcodes could potentially lead to nearly ambiguity-free connectomes via informatic deconvolution at the synapses.

**Supplementary Figure 3** shows a simulation of these statistics for small $N$ and comparison with the above simple model.

**Figure 3.** Scaling properties of informatic deconvolution from a known barcode pool. Simulated barcode pools were generated randomly from an equal mixture of $A$, $T$, $C$ and $G$ (with a check to ensure uniqueness within the pool) and the probability of ambiguous informatic deconvolution was evaluated as a function of the barcode length $N$ and barcode pool size $n$. Solid lines: simulations. Dotted lines: the model $\log_2(P_{\text{ambiguous}}) = 2 \times \log_2(n) + (\log_2(1.75) - 4) \times N - 1$.

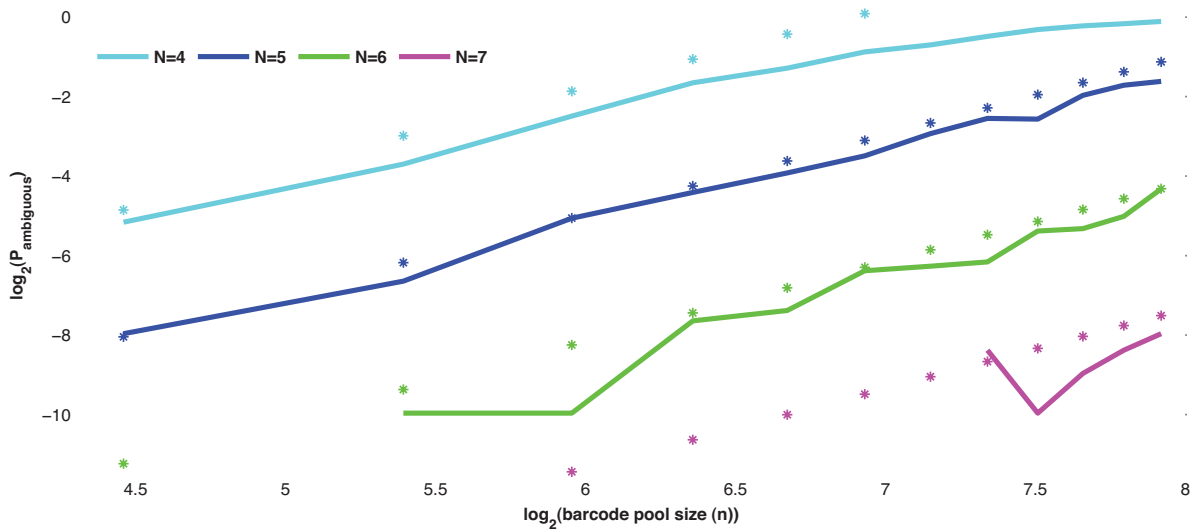

**Programmatic sequencing**     Additional *multi-base* sequencing reactions can also provide the missing 0.75 bits of information. These can be implemented using a sequencing by ligation strategy, which selects for correct hybridization of a multi-base ligation probe. Ligation efficiency is decreased by orders of magnitude by the presence of a single nucleotide mismatch within a symmetrical 12-base footprint, six bases on each side of the single-strand break [91].

As an example, through the appropriate choice of ligation probe-sets, a fluorescent signal could be indicative that bases 1 and 2 of the read are the *same* base (reminiscent of a logical AND gate). If we had previously observed A and T signals at base 1, and A and G signals at base 2, this additional information would unambiguously imply that the two barcodes begin with the sequences "AA" and "TG" (the alternative possibility "AG" and "TA" is ruled out because the extra multi-base read has indicated that the first two bases are identical in at least one of the molecules). Such "parity" sequencing reactions must be programmed, either through the design of the template molecules, or through the enzyme and sequencing reagents interrogating the template molecules. This scheme would increase the number of required biochemical cycles.

**Using informatic deconvolution for FISSEQ error-correction**     FISSEQ inside the cell's nucleus can "average" over many rolonies corresponding to the same cell-barcode, allowing unambiguous knowledge of the *pool* of all cell-barcodes (not to mention their corresponding soma positions) via nuclear FISSEQ. This knowledge could be used – in a manner analogous to informatic deconvolution – to disambiguate error-prone sequencing reads from the synapses, which might derive from only 1-2 rolonies. This could be valuable, for example, for "factoring out" the error rate associated with the reverse transcriptase step in RNA FISSEQ. Note that, unlike in traditional bulk sequencing library preparation, FISSEQ does not involve sequential, bottlenecked amplification steps that can, e.g., "lock in" polymerase errors generated during the first few cycles of PCR; indeed, in rolling-circle amplification, only the original template molecule itself is copied.

## 7.5  Hybrid strategies

Hybrid strategies could combine favorable aspects of BOINC and FISSEQ-BOINC. One possibility is as follows: 1) first localize each neuron and identify its corresponding barcode via FISSEQ of the nuclei, and then 2) subsequently subject the entire brain, post-FISSEQ, to bulk sequencing to determine the barcode *pairs* via BOINC [3–5]. In this scenario, only (multiple redundant copies of) a single barcode – the self-barcode of a particular cell – is in situ sequenced in each nucleus, greatly reducing the optical resolution requirement and hence allowing a significant speedup.

**Cost estimate**: To image only the nuclei, which are separated by $(1\,\mathrm{mm}^3/100000)^{1/3} \approx 20\,\mu\mathrm{m}$ on average, a wide-field scan could first be applied to locate the nuclei. Then, using random-access confocal scanning in 3D, at 10 ms dwell time per pre-identified nucleus (a somewhat arbitrary estimate), the per-cycle imaging time would be $10\,\mathrm{ms} \times \rho \times 420\,\mathrm{mm}^3$ where $\rho = 100\,000/\mathrm{mm}^3$ is the approximate density of neuronal nuclei. Thirty cycles of FISSEQ on all mouse brain neuronal nuclei would then take only five months on a single machine, i.e., for a cost in the range of \$200k. The tissue could then be subjected to bulk sequencing of barcode *pairs* corresponding to synaptically connected cells, at the cost associated with BOINC, i.e., potentially <\$1M [5]. The hybrid process could thus achieve a 1-year project cost in the \$1M range for a whole mouse brain barcode-based connectome with precisely known soma position for each barcoded cell.